\documentclass[12pt]{iopart}

\usepackage{graphicx}% Include figure files
\usepackage{iopams}  

\usepackage{colortbl}
\definecolor{Gray}{gray}{0.85}
\newcolumntype{a}{>{\columncolor{Gray}}c}

\begin{document}

%\end{document}

\title{Contrasting Freezeouts in Large Versus Small Systems}
\author{Sandeep Chatterjee, Ajay Kumar Dash and Bedangadas Mohanty}
\address{School of Physical Sciences,\\ National Institute 
of Science Education and Research, Jatni-752050, India}
\ead{sandeepc@niser.ac.in, ajayd@niser.ac.in and bedanga@niser.ac.in}

\begin{abstract}
We study the data on mean hadron yields and contrast the chemical freezeout conditions in 
p+p, p+Pb and Pb+Pb at the Large Hadron Collider (LHC) energies. We study several schemes 
for freezeout that mainly differ in the way strangeness is treated: i. strangeness freezes out 
along with the non-strange hadrons in complete equilibrium (1CFO), ii. strangeness freezes out 
along with non-strange hadrons with an additional parameter $\gamma_S$ accounting for non-equilibrium 
production of strangeness (1CFO+$\gamma_S$), and iii. strangeness freezes out earlier than non-strange 
hadrons and in thermal equilibrium (2CFO). A comparison of the chisquares of the fits indicate a 
dependence of the freezeout scheme on the system size. The minimum bias p+p and different centralities 
of p+Pb and peripheral Pb+Pb data prefer 1CFO$+\gamma_S$ with $\gamma_S$ approaching unity as we go 
from p+p to central p+Pb and peripheral Pb+Pb. On the other hand, the mid-central to central Pb+Pb 
data prefer 2CFO over 1CFO+$\gamma_S$. Such system size dependence of freezeout scheme could be an 
indication of the additional interaction in Pb+Pb over p+Pb and p+p.

%
%\PACS{
%      {PACS-key}{discribing text of that key}   \and
%     {PACS-key}{discribing text of that key}
%     } % end of PACS codes
\end{abstract}
\maketitle

\section{Introduction}
A hot and dense strongly interacting medium, the fireball, is produced in ultra-relativistic 
heavy ion collisions (HICs). Assuming rapid thermalization, the ensuing expansion phase has 
been modelled by relativistic hydrodynamics with considerable success~\cite{hydro}. The expansion 
of the fireball causes dilution of the energy density of the system. The medium mean free path 
elongates and finally drives the fireball out of equilibrium. The analysis of the hadron yields 
allows an access to the properties of the chemical freezeout (CFO) surface where the hadronic 
inelastic collisions ceased~\cite{cfo1,cfo2,cfo3,cfo4}. There are on going efforts to extract 
the CFO surface not only from the measured mean hadron yields but also from their higher order 
moments through the measurement of event by event (e-by-e) fluctuations of the QCD conserved 
charges~\cite{fluc1,fluc2,fluc3,fluchrg,fluc4,fluc5}.\\

A unified CFO scheme where all the hadrons freezeout together, parametrised 
by a single temperature, baryon chemical potential and volume of the freezeout surface (1CFO) 
have been widely used to describe the hadron yields across a wide 
range of $\sNN$ ~\cite{cfo1,cfo2,cfo3,cfo4}. The extracted CFO surface from these analyses was found in close 
proximity to the hadronization surface as indicated by lattice QCD studies at zero net-baryon 
densities~\cite{LQCDhad}. This led to several suggestions like the hadrons are born directly into 
equilibrium at the time of hadronization~\cite{stock} or the multi-particle interactions that 
maintain equilibrium are relevant only close to the hadronization surface~\cite{multi-particle}. 
However, the absence of an unambiguous microscopic description of the freeze-out dynamics and 
that the interaction cross-sections among various hadrons in the system not playing a manifestly 
important role left a sense of incompleteness in the 1CFO approach. In addition, it was observed 
that this approach is reasonably well applicable to $e^++e^-$, p+p and A+A 
collisions~\cite{BecattinieeAA}, raising questions as to the sensitivity of the 1CFO scheme to 
the differences in the underlying interactions as expected for such a variety of systems.

It is now well known that the standard 1CFO scheme fails to describe satisfactorily the hadron yields 
for Pb+Pb collisions at $\sNN=2.76$ TeV~\cite{LHC} which was termed as `proton anomaly'~\cite{panomaly} 
and also interpreted as tension in the non-strange vs strange sectors~\cite{Preghenella,Bellwied2trans,
2cfothms,floris,2cfonuclei}. It is to be noted that similar tensions between 1CFO and 
data exist even at lower $\sNN$~\cite{BecattiniSPS,2cfothms,2cfonuclei,VovchenkoSPS}.
There have been suggestions to describe the hadron yields at the LHC- incorporating 
effects of non-equilibrium evolution either through a microscopic hadronic 
afterburner~\cite{Steinurqmd, Becattiniurqmd} or through additional flavor dependent fugacity 
factors~\cite{noneq1,noneq2,noneq3} within the 1CFO scheme. It has been further suggested that the current 
discrepancy between models and data could be due to missing resonances in the models that have not been 
observed yet but are expected from QCD models and LQCD computations~\cite{missingStrange,Hagedorn}. A 
flavor dependent sequential freezeout scheme with different freezeout surfaces for hadrons 
with zero and non-zero strangeness content (2CFO) was also suggested based on flavor dependence in 
several thermodynamic quantities in LQCD computations~\cite{Bellwied2trans} and known systematics of 
hadron cross-sections~\cite{2cfothms}. The 2CFO scheme was analysed and was found to provide a much 
improved description of the hadron yields~\cite{2cfothms, 2cfobugaev, 2cfonuclei, 2cfospectra}. 
An alternate approach that has successfully described the LHC yields is to consider 
the Van der Waals excluded volumes for the hadrons with the hadron mass-eigenvolume relationship 
being flavor dependent~\cite{AlbaVolume} (2EV). The $\chi^2/\textrm{NDF}$ (NDF=Number of data points - 
Number of free parameters) for the data-model comparison to 
the LHC hadron yields, for both the schemes, 2CFO and 2EV are less than unity. Thus, it is not possible to 
discriminate between these two very different approaches to introduce flavor dependence with yield data 
alone. Hence, it will be interesting to check how 2EV fares with the data on hadron spectra at the LHC 
energies as well as at the lower energies. For both these cases, it has been already demonstrated that 2CFO fares 
better than the conventional 1CFO scheme~\cite{2cfothms, 2cfospectra}.

Recently it was reported that for p+p collisions at the LHC at $\sNN=900$ GeV and 7 TeV, 1CFO$+\gamma_S$ 
performs better than 2CFO~\cite{ppsabita}. In 1CFO$+\gamma_S$, we still have a unified freezeout for 
all hadrons, however strangeness is relaxed to go out of equilibrium compared to the rest. $\gamma_S$ is the 
additional parameter that accounts for non-equilibrium production of strangeness. This probably 
hints at a system size dependence in the freezeout scheme where HICs prefer 2CFO while small systems like 
p+p and p+Pb prefer 1CFO+$\gamma_S$ scheme. In this paper we aim to investigate this possible system size 
dependence in the freezeout scheme in more details. We have analysed the mid-rapidity data for hadron yields in 
p+p~\cite{ppnonstrange,ppphi,ppstrangebar}, p+Pb~\cite{ppbnonstrange,ppbphi,ppbstrangebar} and 
Pb+Pb~\cite{pbpbnonstrange,pbpbphi,pbpblambda,pbpbstrangebar} collisions at $\sNN=$7, 5.02 and 2.76 TeV 
respectively across all the available centralities. In doing so we cover over three orders of magnitudes in 
terms of the mid-rapidity charged multiplicty as well as the extracted fireball volume at the time of freezeout. 
This allows us to systematically study the system size dependence of the freezeout conditions. The p+p data at 
different LHC energies have been previously studied~\cite{ppsabita}. The thermal freezeout parameters are almost 
constant across these beam energies. Hence the difference in $\sNN$ for p+p, p+Pb and Pb+Pb in our study is not 
expected to affect the conclusions on the role of system size. There have been recent studies on system size 
dependence of hadron yields at SPS energies~\cite{VovchenkoSPS} and the fate of the maxima in several particle 
ratios like $K^+/\pi^+$ and $\Lambda/\pi$~\cite{Oeschler}.

\section{Freezeout Schemes}
The primordial yields in the hadron resonance gas model (HRGM) which is used to extract the 
freezeout conditions in the 1CFO scheme in the Grand Canonical ensemble (GCE) is 
given by
\beqa
N_{i} &=& \frac{g_{i}V}{2\pi^{2}}\sum_{k=1}^{\infty}(\pm1)^{k+1}\frac{m_{i}^{2}T}{k} 
 ~K_{2}\left( \frac{km_{i}}{T} \right)e^{k\mu_i/T}.
\label{eq.primmult}
\eeqa
Here, $V$ and $T$ are the volume and temperature of the fireball. $K_2$ is the modified Bessel 
function of the second kind. $g_i$, $m_i$ and $\mu_i$ refer to the degeneracy factor, mass and 
hadron chemical potential of the $i^{th}$ hadron species respectively. The total yield $N_i^{{tot}}$ 
is obtained by including the resonance decay contribution to the above primordial yield
\beqa
N_i^{{tot}} &=& N_i + \sum_j{B.R.}_{ij}\times N_j
\label{eq.totmult}
\eeqa
where ${B.R.}_{ij}$ is the branching ratio for the $j^{th}$ hadron species to $i^{th}$ hadron species. 
As of yet, this model has no predictive power since in principle, all the $\mu_i$s are independent of each 
other. At this stage, a crucial assumption is made of complete chemical equilibrium which allows 
to parametrise all the $\mu_i$s in terms of only $\mu_B$, $\mu_Q$ and $\mu_S$- three chemical 
potentials corresponding to the three conserved charges of QCD, namely baryon number $B$, electric 
charge $Q$ and strangeness $S$
\beqa
\mu_i &=& B_i\mu_B + Q_i\mu_Q + S_i\mu_S
\label{eq.mui}
\eeqa
Now this is a phenomenologically viable model with only $T$, $\mu_B$ and $V$ to fit from data while 
$\mu_S$ and $\mu_Q$ are extracted from the conditions of strangeness neutrality and net baryon to 
net charge ratio of the colliding systems. Thus, 1CFO is expected to work well if all the inelastic 
hadronic interactions freezeout together, in particular, if all the strangeness changing 
transmutations cease at the same instant as those that do not involve changing of strangeness.

In 2CFO, we use different freezeout parameters for hadrons with zero and non-zero 
strangeness content in Eq.~\ref{eq.primmult}. We also employ a variant of 2CFO where we reduce one 
free parameter by imposing entropy conservation between the strange and non-strange freezeout 
surfaces~\cite{2cfobugaev}. The non-strange volume $V_{NS}$ is computed from the entropy conservation 
constraint in the following way
\beqa
V_{NS}s(T_{NS},\{\mu\}_{NS}) &=& V_Ss(T_S,\{\mu\}_S)\label{eq.FS}
\eeqa
where $s$ refers to the entropy density and $\{\mu\}$ just refers to the three chemical potentials 
together: $\mu_B$, $\mu_S$ and $\mu_Q$. We label this scheme as 2CFO+FS. The strange hadrons 
freezeout at the strange freezeout surface. After this, we have only the non-strange hadrons 
following the $(T,\{\mu\})$ trajectory of the fireball. For entropy conservation between the two freezeouts, 
the entropy content of the non-strange hadrons post strange freezeout must be equal to the entropy content 
of the non-strange hadrons prior to the non-strange freezeout. Thus, only the non-strange hadrons contribute 
to both LHS and RHS of Eq.~\ref{eq.FS}. It is to be noted that the strange resonances decay to strange as 
well as non-strange hadrons. Thus the non-strange fitted parameters are affected by the strange hadron yields 
as well.

We have performed our analysis for four different freezeout schemes- 1CFO, 1CFO$+\gamma_S$, 2CFO and 2CFO+FS. 
The hadrons whose mid-rapidity yields have been fitted to extract the CFO conditions are $\pi^++\pi^-$, $K^++K^-$, $p+\bar{p}$, $\phi$, 
$\Xi+\bar{\Xi}$ and $\Omega+\bar{\Omega}$ in all the systems. Additionally, we have also used $\Lambda+\bar{\Lambda}$ 
in p+Pb and only $\Lambda$ in Pb+Pb. At these energies, the data reveals good particle-antiparticle 
symmetry allowing us to take the hadron chemical potentials to be zero throughout this work. Thus the 
parameters to be extracted from the thermal fits in 1CFO are only the fireball volume $V$ and temperature $T$ 
at the time of freezeout. In the 1CFO$+\gamma_S$ scheme, we treat $\gamma_S$ also as an additional fitting parameter. 
In the 2CFO scheme we stay within complete equilibrium setup. However, since we allow for different freezeout 
thermal parameters for the non-strange and strange hadrons we have four ($V_S$, $T_S$, $V_{NS}$ and $T_{NS}$) 
parameters. Finally, in 2CFO+FS, we manage to do away with $V_{NS}$ as a fitting parameter using Eq.~\ref{eq.FS} 
and thus have only three parameters ($V_S$, $T_S$ and $T_{NS}$). Hence, both 1CFO$+\gamma_S$ and 2CFO+FS have 
three free parameters to be fitted from data. Thus, their goodness of fit can be compared unambiguously 
using the chisquares of the fits. 

It has been a standard practice to treat the QCD conserved charges in HICs in the GCE while those of p+p in either 
canonical ensembles for all the conserved charges (CE)~\cite{CE} or strangeness alone (SCE)~\cite{SCE}. In QCD, 
particle production has to obey local charge conservation rules for $B$, $Q$ and $S$. Thus, an independent 
particle production scenario as envisaged within a GCE should receive corrections due to mutli-particle correlations 
introduced through such conservation rules. These corrections are particularly significant in small systems where 
the total multiplicity itself is small. This calls for different ensembles like CE or SCE in small systems. However, 
in a QGP state the flavor carrying degrees of freedom are almost massless, allowing for a profuse productions of 
all the flavors. In such cases, the corrections due to the conservation laws will be small and GCE can provide a 
good description of particle production even in small systems. Also, at high energies when we only treat the 
mid-rapidity data, it might be possible to treat the data from small systems like p+p and p+Pb as well in the 
GCE as we are always observing an open system irrespective of the system size. In a recent work it was pointed 
out that at the top RHIC and LHC energies, the mid-rapidity p+p data can be better described in the GCE than in 
CE or SCE~\cite{ppsabita}. We have checked that the $\chi^2/\textrm{NDF}\sim20$ in the CE and SCE while in GCE we 
find $\chi^2/\textrm{NDF}\sim3$. In this paper, we always work in the GCE across all the system sizes in p+p, p+Pb 
and Pb+Pb.

\section{Freezeout in p+p, p+Pb and Pb+Pb}

\begin{table} [ht]
\begin{center}
\caption{The goodness of fits of different models compared. The number of free parameters 
are 2, 3, 4 and 3 for 1CFO, 1CFO$+\gamma_S$, 2CFO and 2CFO+FS respectively. The number of 
yield data available for p+p, p+Pb and Pb+Pb are 6, 7 and 7 respectively. Since 1CFO$+\gamma_S$ 
and 2CFO+FS have same number of free parameters, their goodness of fit can be compared 
unambiguously by simply comparing the chisquares of their fits to data. These columns have been 
shaded in grey. We observe that 1CFO$+\gamma_S$ has smaller chisquare for p+p and p+Pb collisions 
while 2CFO+FS perform better in Pb+Pb collisions. This points to a plausible system size dependence 
of freezeout scheme. \\}
\label{tab.chisq}
\resizebox{\textwidth}{!}{%
\begin{tabular}{|c|c|c|a|c|a||c|c|c|c|} 
\hline
 System & Centrality $\%$ & \multicolumn{4}{|c||}{$\chi^2$} & \multicolumn{4}{|c|}{$\chi^2/\textrm{NDF}$}\\\cline{3-10}
 & & 1CFO & 1CFO$+\gamma_S$ & 2CFO & 2CFO+FS & 1CFO & 1CFO$+\gamma_S$ & 2CFO & 2CFO+FS\\\hline\hline
 p-p & min-bias & 62.6 & 14.2 & 24.2 & 27 & 15.7 & 4.7 & 12.1 & 9 \\\hline\hline
 p-Pb & 00-05 & 18.67 & 16.72 & 17.56 & 18.36 & 3.7 & 4.2 & 5.9 & 4.6 \\\cline{2-10}
      & 05-10 & 20.06 & 17.08 & 17.35 & 19.90 & 4 & 4.3 & 5.8 & 5 \\\cline{2-10}
      & 10-20 & 21.44 & 16.42 & 19.38 & 21.40 & 4.3 & 4.1 & 6.5 & 5.4 \\\cline{2-10}
      & 20-40 & 24.69 & 18.17 & 22    & 24.40 & 5 & 4.5 & 7.3 & 6.1 \\\cline{2-10}
      & 40-60 & 32.43 & 17.35 & 26.74 & 29.27 & 6.5 & 4.3 & 8.9 & 7.3 \\\cline{2-10}
      & 60-80 & 39.35 & 14.58 & 26.91 & 31.50 & 7.9 & 3.6 & 9 & 7.9 \\\cline{2-10}
      & 80-100 & 39.07 & 10.61 & 21.46 & 28 & 7.8 & 2.7 & 7.2 & 7 \\\hline\hline
 Pb-Pb & 00-10 & 17.9 & 10.59 & 2.39 & 3.85 & 3.6 & 2.6 & 0.5 & 1 \\\cline{2-10}
       & 10-20 & 22.38 & 12.22 & 1.52 & 3.27 & 4.5 & 3.1 & 0.5 & 0.8 \\\cline{2-10}
       & 20-40 & 27.01 & 12.57 & 1.03 & 2.50 & 5.4 & 3.1 & 0.3 & 0.6 \\\cline{2-10}
       & 40-60 & 14.38 & 9.35 & 2.01 & 3.57 & 2.9 & 2.3 & 0.7 & 0.9 \\\cline{2-10}
       & 60-80 & 9.95 & 8.28 & 7.6 & 9.90 & 2 & 2.1 & 2.5 & 2.5 \\\hline\hline
\end{tabular}%
}
\end{center}
\end{table}

\begin{figure}
 \begin{center}
 \includegraphics[scale=0.36]{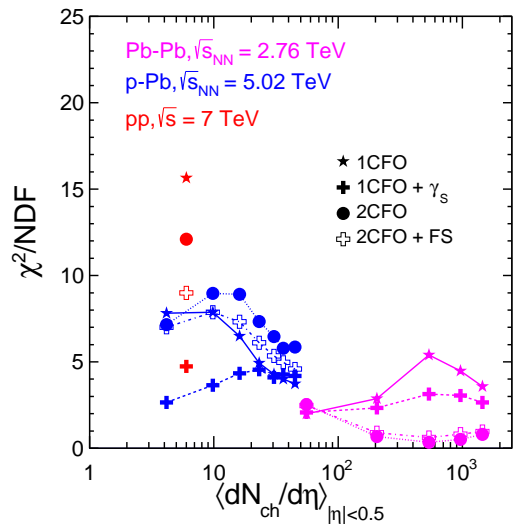}
 \caption{(Color online) The $\chi^2/\textrm{NDF}$ for the different freezeout schemes as 
 tabulated in Table~\ref{tab.chisq} have been plotted. It is interesting to note that while 
 2CFO and its variant 2CFO+FS perform better for $dN_{ch}/d\eta>\sim10^2$, 1CFO$+\gamma_S$ 
 seems to be preferred for lower $dN_{ch}/d\eta$.}
 \label{fig.chisq}
 \end{center}
\end{figure}

We first compare the quality of fit in different schemes and systems. The goodness of fit in terms of 
$\chi^2/\textrm{NDF}$ ($\textrm{NDF}=$Number of data points - Number of free parameters) for the different 
freezeout schemes across different centralities for Pb+Pb, p+Pb and p+p collisions are tabulated in 
Table~\ref{tab.chisq} and plotted in Fig.~\ref{fig.chisq}. In all the schemes the fit quality deteriorates 
as we go from Pb+Pb to p+Pb and p+p collisions. This is most dramatic for the 2CFO and 2CFO+FS schemes 
where the $\chi^2/\textrm{NDF}$ rises from $\sim0-2$ in Pb+Pb to $\sim6-9$ in p+Pb and p+p. In 1CFO, 
$\chi^2/\textrm{NDF}$ ranges between $2-5$ in Pb+Pb while in p+Pb and p+p it varies between $4-8$. 
1CFO$+\gamma_S$ scheme gives largely a constant $\chi^2/\textrm{NDF}$ with a very mild rise from Pb+Pb 
($\chi^2/\textrm{NDF}$ $\sim2-3$) to p+Pb and p+p ($\chi^2/\textrm{NDF}$ $\sim3-4$).

The most important lesson that we learn from Fig.~\ref{fig.chisq} is that the 2CFO 
freezeout scheme works only for large system size. In large systems, the fireball lifetime 
is expected to be long resulting in sufficient interaction amongst constituents and 
thus the role of hadron-hadron cross sections come into play. The kaons mainly maintain 
the chemical equilibrium between the non-strange and strange sectors. In a cooling fireball, the kaon 
to pion ratio rapidly falls resulting in an early freezeout for strangeness~\cite{2cfothms}. This is 
the microscopic picture that is the motivation behind the 2CFO scheme. Thus, 
high rate of hadronic interactions resulting in a hadronic medium formation is essential 
for 2CFO to work. On the other hand, in small systems the fireball lifetime is much shorter as 
the expansion dynamics is dominant resulting in little hadronic interactions. Thus the 
hierarchy in hadron-hadron cross-sections do not enter the freezeout dynamics and we have 
a sudden freezeout scenario with all the hadrons freezing out together. 

In order to understand whether the better performance of 2CFO over 1CFO$+\gamma_S$ is 
merely due to more free parameters, we have also performed the fits in 2CFO+FS which has 
the same number of free parameters as 1CFO$+\gamma_S$ and hence any difference in the chisquares of 
2CFO+FS and 1CFO$+\gamma_S$ is solely due to the difference in the freezeout scheme of strangeness. 
As seen in Table~\ref{tab.chisq} and Fig.~\ref{fig.chisq}, the performance of 2CFO+FS is better than 
1CFO$+\gamma_S$ in Pb+Pb while the latter performs better in p+Pb and p+p collisions. This strongly 
suggests a system size dependence in the freezeout scheme of strangeness. Since 2CFO and 2CFO+FS 
provide similar results of fits to hadron yield data, henceforth we will only show the results from
2CFO.

\begin{figure}[t]
 \begin{center}
 \includegraphics[scale=0.36]{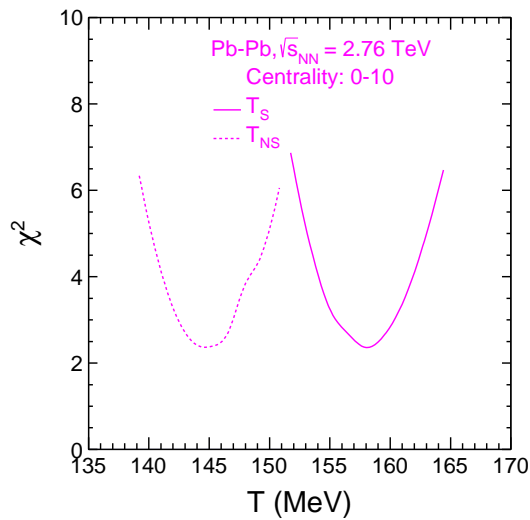}
 \caption{(Color online) The $\chi^2$ vs $T$ for the non-strange and strange freezeout 
 surfaces in 2CFO.}
 \label{fig.chisqvsT}
 \end{center}
\end{figure}

We will now investigate the stability of our fits in 2CFO. Pions and protons are the only 
hadrons with zero strangeness content whose yields are available in the experiments. This might render 
the extraction of two parameters, $T_{NS}$ and $V_{NS}$ unstable from the fits to data. However, as mentioned 
earlier, the heavier strange resonances decay to strange as well as non-strange hadrons. This couples the 
non-strange freezeout parameters to the strange hadron yields as well and allows for a reliable extraction of the 
freezeout parameters. As seen in Fig.~\ref{fig.chisqvsT}, the $\chi^2$ vs $T$ profile for the 
non-strange and strange freezeout surfaces exhibit a narrow parabola like structure with their minima well 
separated indicating a clear hint of separate freezeouts of the non-strange and strange hadrons.

\begin{figure*}[t]
 \begin{center}
 \includegraphics[scale=0.8]{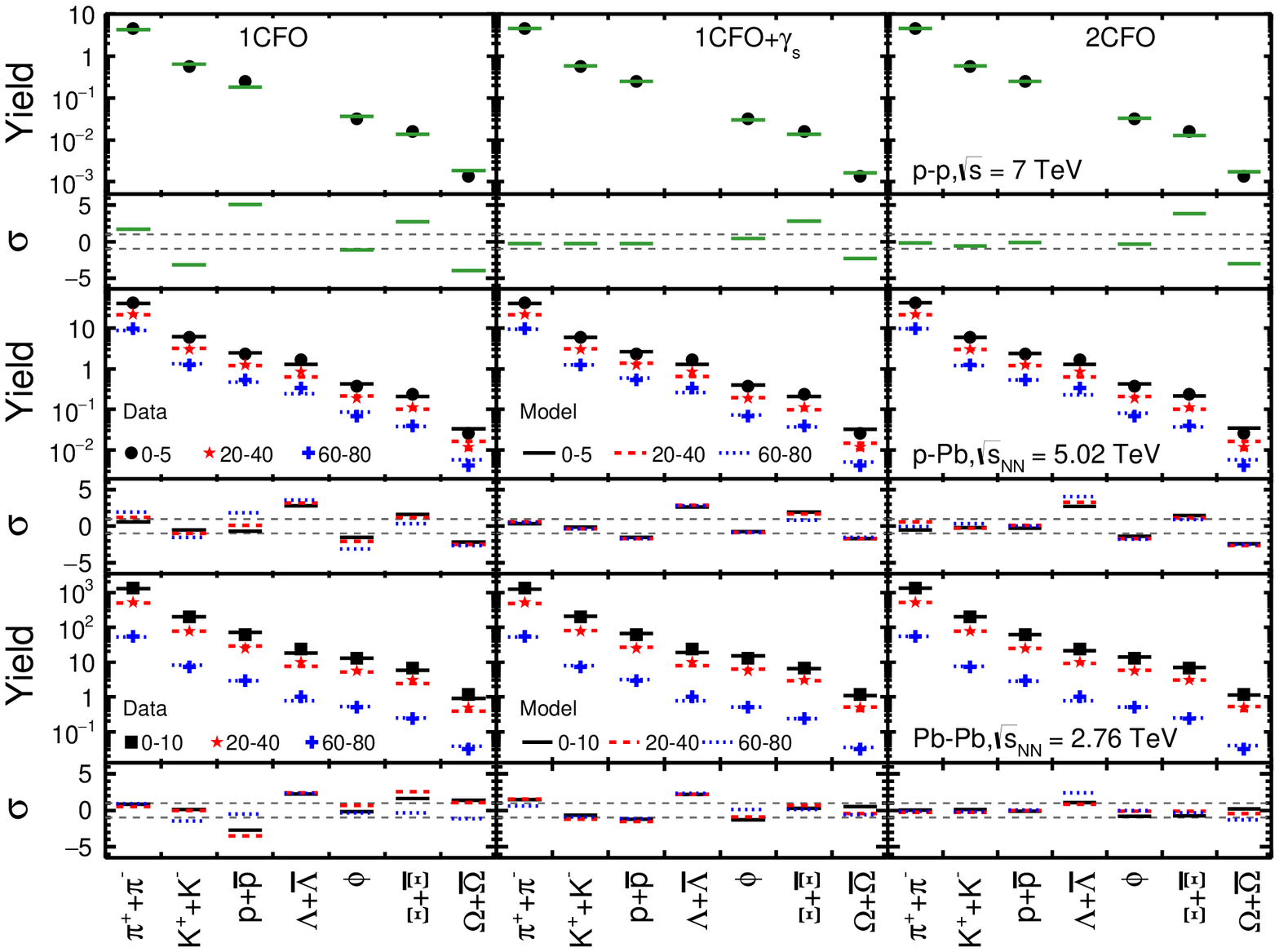}
 \caption{(Color online) The comparison between data~\cite{ppnonstrange,ppphi,ppstrangebar,
 ppbnonstrange,ppbphi,ppbstrangebar,pbpbnonstrange,pbpbphi,pbpblambda,pbpbstrangebar} and model 
 for p+p (top row)~\cite{ppnonstrange,ppphi,ppstrangebar}, p+Pb (middle row)~\cite{ppbnonstrange,
 ppbphi,ppbstrangebar} and Pb+Pb (bottom row)~\cite{pbpbnonstrange,pbpbphi,pbpblambda,pbpbstrangebar} 
 systems in three different freezeout schemes: 1CFO (left column), 1CFO$+\gamma_S$ (middle column) and 
 2CFO (right column). Note that for Pb+Pb the $\Lambda+\bar{\Lambda}$ column refer to only $\Lambda$ as 
 data for $\bar{\Lambda}$ is not available. Also shown are the deviation of the data from the model 
 (Eq.~\ref{eq.dev}) for each freezeout scheme.}
 \label{fig.dev}
 \end{center}
\end{figure*}

We have plotted the measured hadron yields in min-bias p+p~\cite{ppnonstrange,ppphi,ppstrangebar} and different 
centralities of p+Pb~\cite{ppbnonstrange,ppbphi,ppbstrangebar} and Pb+Pb~\cite{pbpbnonstrange,pbpbphi,pbpblambda,
pbpbstrangebar} collisions and compared with the thermal model fits for the three freezeout schemes in Fig.~\ref{fig.dev}. 
For clarity, we show results for only three centralities in p+Pb and Pb+Pb, representing central, mid-central 
and peripheral collisions. In the lower panel of each figure, we have also shown the deviation $\sigma$ given by
\beqa
\sigma&=&\frac{{\mathrm{Data - Model}}}{{\textrm{Error of Data}}}.
\label{eq.dev}
\eeqa
In p+p, the 1CFO scheme fares poorly with $\sigma>5$ for proton. The 1CFO$+\gamma_S$ and 2CFO schemes seem to 
fare equally good, however the multi-strange baryons $\Xi$ and $\Omega$ have $\sigma>2$. Similar situation as 
in p+p prevails through peripheral to mid-central p+Pb collisions. There is considerable tension both in the 
mesonic as well as baryonic sectors in the 1CFO scheme which improves as we include $\gamma_S$ in the 1CFO$+\gamma_S$ 
scheme.  The situation changes slightly for the central bins. Here the non-strange sector fares much better in the 
1CFO scheme while the $\sigma\sim2$ in the strange sector do not go away with more free parameters in the 
1CFO$+\gamma_S$ and 2CFO schemes. In the 2CFO scheme with one more free parameter the agreement for proton betters 
but at the cost of $\phi$ mesons. Overall we don't gain by including an extra parameter in 2CFO as compared 
to 1CFO$+\gamma_S$. 

We now turn our attention to the bottom row of Fig.~\ref{fig.dev} where we have shown results 
for the Pb+Pb case. The scenario for the most peripheral Pb+Pb bin is 
similar to the central p+Pb bins. All the schemes exhibit similar goodness of fit and the gain 
in improving the fit marginally in going from 1CFO to other schemes is compensated by the 
additional free parameters required. The other centralities exhibit strong tension between 
the non-strange and strange baryons in 1CFO. In 1CFO$+\gamma_S$, the agreement for the 
strange baryons improve but the tension between proton and $\Lambda$ continues while a mild 
tension between $\pi$ and kaons develop. Unlike in p+p and p+Pb, we find an appreciable 
improvement in the fits in 2CFO, only $\Lambda$ shows a mild deviation of $\sigma\sim1$.

\begin{figure}
 \begin{center}
 \includegraphics[scale=0.36]{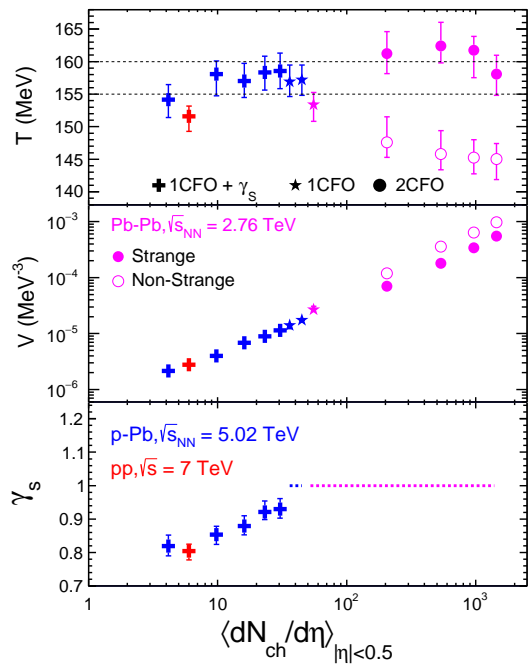}
 \caption{(Color online) The best fitted freezeout parameters (temperature $T$, volume $V$ and 
 strangeness non-equilibrium factor $\gamma_S$) amongst the three different 
 freezeout schemes. }
 \label{fig.bestparameter}
 \end{center}
\end{figure}

In Fig.~\ref{fig.bestparameter} we have plotted the freezeout parameters that provide the best 
description to the data amongst the three different freezeout schemes that we have studied here.  
The freezeout temperature stays around $155-160$ MeV for p+p, p+Pb and also for the 
strange hadrons in Pb+Pb collisions. However, the non-strange hadrons in Pb+Pb breakaway from this 
trend. The higher rate of interaction in the strongly interacting hadronic medium casues a delayed 
freezeout for them and they finally freezeout at temperatures which are lower by about $10\%$ than 
those of the strange freezeout surface. All the centralities except 
the most peripheral bin of $(60-80)\%$ centrality in Pb+Pb is best described by 2CFO. The non strange 
freezeout temperature monotonically rises as one goes from central to peripheral bins and merge with its 
strange companion in the $(60-80)\%$ centrality bin. The $(60-80)\%$ centrality bin in Pb+Pb and the two 
most central bins of $(0-5)\%$ and $(5-10)\%$ of p+Pb are best described by the 1CFO scheme. For the 
rest of the centralities in p+Pb and p+p, it is essential to fit $\gamma_S$. Within the 1CFO$+\gamma_S$ 
scheme, the extracted $\gamma_S$ in p+Pb varies from $0.95-0.80$ for $(10-20)\%$ to $(80-100)\%$ 
centrality bins. We note here that unlike in the Pb+Pb case, this is more in accordance with expectation 
from the core-corona picture where the decreasing trend of $\gamma_S$ can be understood as the 
decreasing contribution of the core of the colliding Pb ion as one goes from central to peripheral 
p+Pb bins. The p+p freezeout conditions are in agreement with that of peripheral p+Pb.

The temperatures extracted in HICs can be contrasted to those obtained using 
the e-by-e fluctuation measures and the hadronisation temperature computed by Lattice QCD. The 
freeze-out temperature obtained from analysis of fluctuation measure involving dominantly non-strange 
hadrons for Au+Au collisions at top RHIC energy is $\sim144\pm6$ MeV~\cite{fluc3, fluchrg} and is in agreement 
with the freeze-out temperature from 2CFO for non-strange hadrons in Pb+Pb collisions 
reported here.  The hadronization temperatures reported by LQCD calculations at zero $\mu_B$ in 
Ref.~\cite{LQCDhad} using strange-quark
number susceptibilities are found to be closer to the freeze-out temperature reported here for 
strange hadrons, while hadronization temperature calculated using chiral condensates gives a value 
which is closer to the freeze-out temperature reported here for non-strange hadrons. However it may 
be noted that the systematic errors associated with the LCQD results are quite large. 

\section{Summary and Outlook}
We have studied the system size dependence in the CFO conditions in 
p+p, p+Pb and Pb+Pb collisions at the LHC. Our study confirms 
that the analysis of the hadron yields within thermal models is sensitive to the physics of the CFO 
and the $\chi^2/\textrm{NDF}$ is a good measure to discriminate between the different CFO scenarios. 
If the analysis technique was blind to the underlying physics of the different CFO scenarios, one 
would expect 2CFO to perform the best across all system sizes as it has the highest number 
of free parameters amongst all the three CFO scenarios studied here. However, we explicitly show that 
this is not the case. 2CFO performs best only in large system sizes with $dN_{ch}/d\eta>100$. For smaller 
systems, 2CFO clearly performs worse than the other scenarios. The varying interplay of expansion and 
interaction amongst the fireball constituents in Pb+Pb versus p+Pb and p+p could give rise to the above 
system size dependence in freezeout.

\section{Acknowledgement}
SC acknowledges discussions and collaborations on freezeout with Rohini Godbole 
and Sourendu Gupta. SC and AKD thank XIIth plan project no. 12-R$\&$D-NIS-5.11-0300 of 
Govt. of India for support. BM acknowledges support from DST SwarnaJayanti and 
DAE-SRC projects of Govt. of India.

\section*{References}	

%
% BibTeX users please use
% \bibliographystyle{}
% \bibliography{}
%
% Non-BibTeX users please use

\end{document}